# TEVATRON COLLIDER OPERATIONS AND PLANS

PETER H. GARBINCIUS
*Accelerator Division*
*Fermi National Accelerator Laboratory*
*Batavia, IL 60510 U.S.A.*

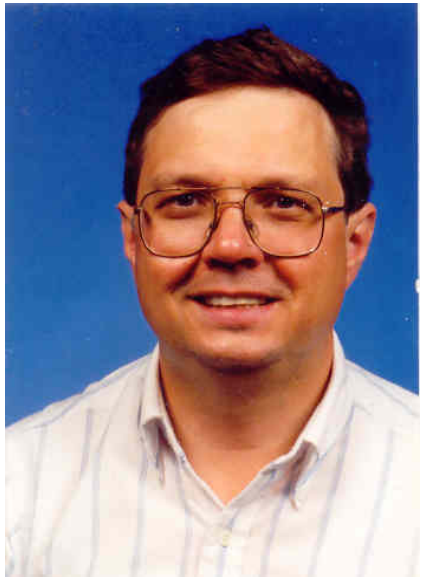

Run II of the Tevatron Collider is reviewed, emphasizing operations through March 15, 2004. The Run II Luminosity Upgrade plans and luminosity projections through 2009 are discussed.

**Introduction**

Fermilab's Tevatron is a proton-antiproton collider with center of mass energy of 1.96 TeV. The antiprotons are produced by 125 GeV protons from the Main Injector striking a stainless steel target. The 8 GeV antiprotons are collected and cooled in the Debuncher and Accumulator rings of the Antiproton Source and, just recently, in the Recycler ring before acceleration by the Main Injector and the Tevatron. In addition to energy, a vital parameter for generating physics data is the Luminosity delivered to the experiments, given by the formula [1]

$$\mathsf{L} = \frac{3\gamma f_0}{\beta^*}(BN_p)\left(\frac{N_{\bar{p}}}{\varepsilon_p}\right)\frac{F(\beta^*,\theta_{x,y},\varepsilon_{p,\bar{p}},\sigma^L_{p,\bar{p}})}{(1+\varepsilon_{\bar{p}}/\varepsilon_p)}$$

where

- $\gamma$ is the relativistic boost (1045 for Tevatron),
- $f_o$ is the revolution frequency (47,700 per second),
- $\beta^*$ is the lattice function at the interaction points (0.35 meter),
- $N_{pbar}$ and $N_p$ are the average number of antiprotons and protons per bunch,
- $B$ is the number of proton and antiproton bunches (36 each),
- $\varepsilon_p$ and $\varepsilon_{pbar}$ are the 95% normalized emittances (~20 $\pi$ x $10^{-6}$ meter),
- $F(\beta^*,\theta_{x,y},\varepsilon_p,\varepsilon_{pbar},\sigma^L_{p,pbar})$ is a geometrical form factor which depends on the transverse sizes of the proton and antiproton beams, the bunch lengths $\sigma^L_{p,pbar}$ (0.5 meter), and the crossing angle $\theta_{x,y}$ (= 0 degrees) at the intersection region ($F$ is typically ~ 0.7).

The most critical feature in delivering high luminosities is the total number of antiprotons stored, $BN_{pbar}$. In principle, increasing the number of protons should increase the luminosity, but the proton beam brightness, $N_p/\varepsilon_p$, is limited by the resulting beam-beam tune shift on the antiproton beam. Therefore, a major thrust of the Run II Luminosity Upgrade program is to maximize the number of antiprotons brought into collision.

The principle elements of the luminosity upgrade plan include: increasing the antiproton production rate by increasing (x2) the number of protons on the production target by the technique of slip-stacking, increasing (x2) the antiproton collection efficiency, increasing (x3) the rate capability to accumulate and to cool antiprotons by upgrading the cooling systems and using the Recycler ring as an additional stage for antiproton collection, and by upgrading the Tevatron to efficiently handle the higher intensity proton and antiproton bunches. These upgrades are discussed in detail in the program plan - The Run II Luminosity Upgrade at the Fermilab Tevatron [2].

**Run II Collider Performance**

The Tevatron Collider was originally designed in the mid-1980's for maximum instantaneous (start of store) luminosities of $L_0^{max}$ = 1 x $10^{30}$ cm$^{-2}$ sec$^{-1}$ and provides collisions simultaneously for the CDF and D0 experiments. During the period August 1992 through March 1996, Run I integrated $\int L\, dt$ = 180 pb$^{-1}$ with $L_0^{max}$ = 25 x $10^{30}$ cm$^{-2}$ sec$^{-1}$, 25 times the original design, during which the data for the discovery of the top quark were taken.

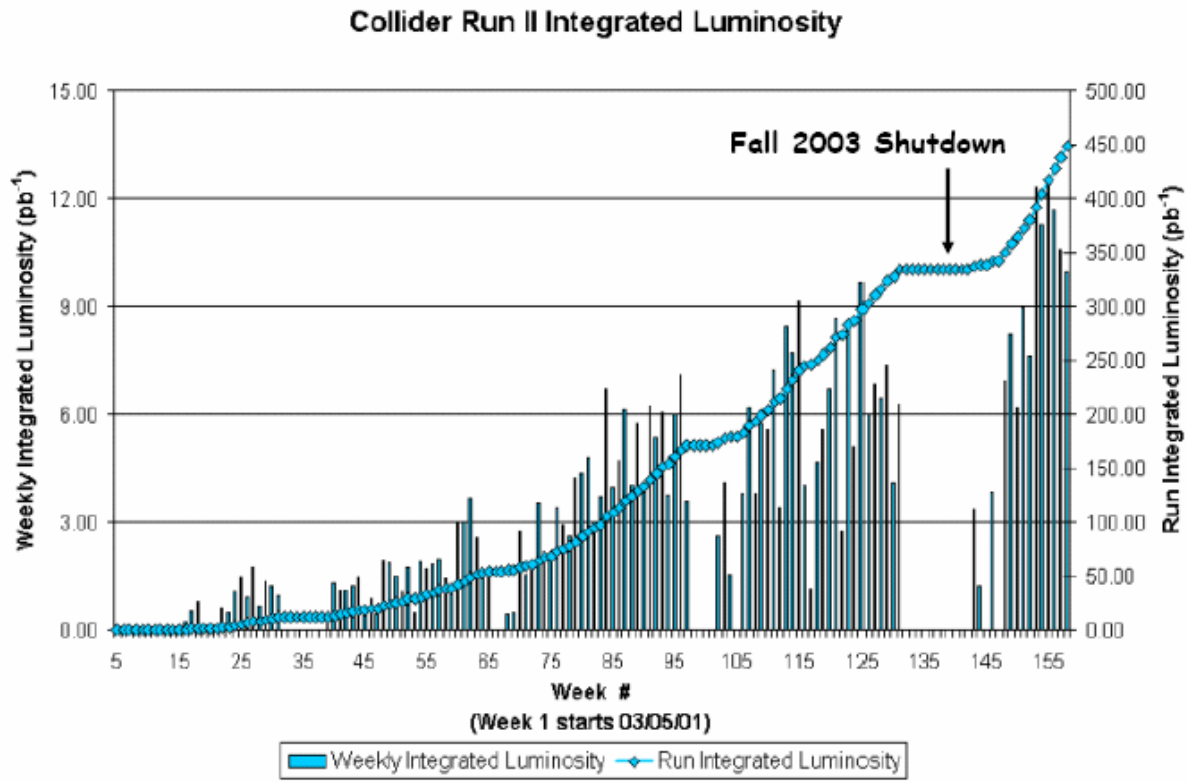

Run II began integrating luminosity in March 2001 and through March 14, 2004 delivered $\int L\, dt$ = 444 pb$^{-1}$ with $L_0^{max}$ = 67 x $10^{30}$ cm$^{-2}$ sec$^{-1}$. In FY2003, the Tevatron delivered 226 pb$^{-1}$, exceeding its Department of Energy goal. Recently, during 2004, the *average* $L_0^{max}$ = 49 x $10^{30}$ cm$^{-2}$ sec$^{-1}$. Run II is planned to continue until August 2009 with projected ranges of $\int L\, dt$ from 4.4 to 8.5 fb$^{-1}$ and $L_0^{max}$ from 160 to 270 x $10^{30}$ cm$^{-2}$ sec$^{-1}$.

(See discussion of luminosity projections below.) The integrated luminosities, by week and total for the entire Run II, are shown on the plot above [3].

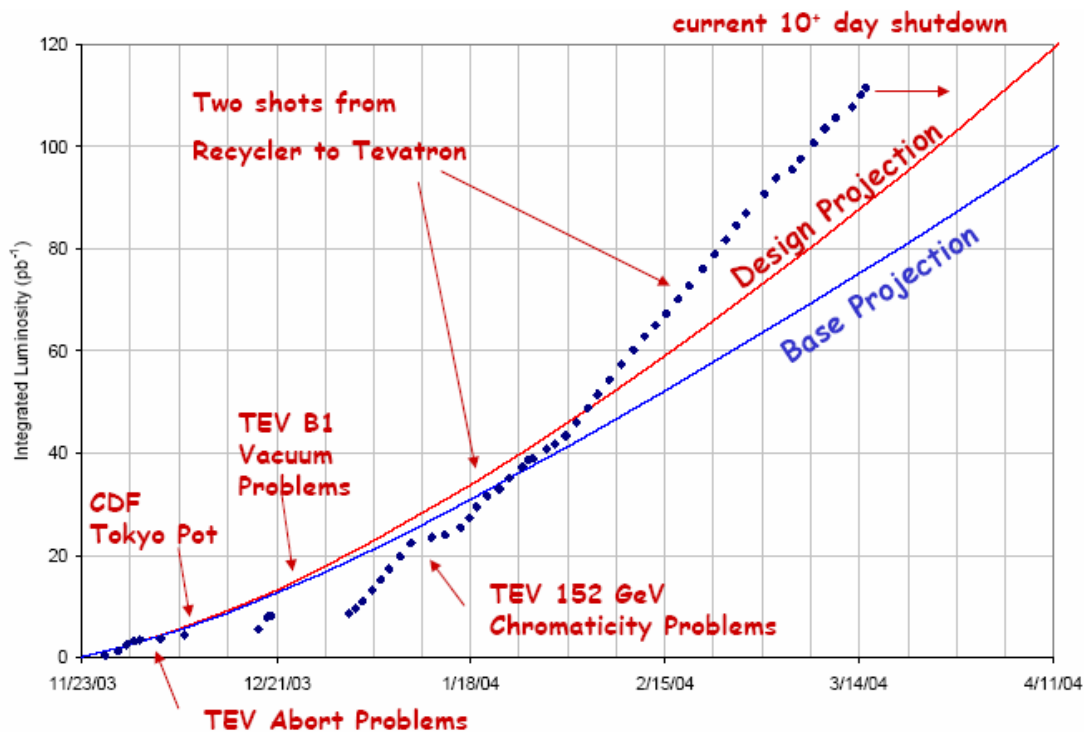

The above plot shows the improved luminosity performance in Fiscal Year 2004 (starting October 1, 2003) after the shutdown in the Fall of 2003. During this 10 week shutdown, the main objective was to improve the vacuum of the Recycler ring. The beam lifetimes and emittance growth rates for the Recycler were greatly improved, surpassing expectations. The stochastic cooling in the recycler is now adequate to support the integration of electron cooling into collider operations. In early 2004, the first two shots of antiproton beams were delivered from the Recycler and produced useable luminosity. The major Tevatron efforts during this shutdown included adjusting alignment of components, adjusting the cryogenic suspensions of dipoles to reduce skew-sextupole components, and inserting a smooth metal liner to reduce the beam impedance of the injection Lambertson magnet. A longitudinal damper system was added to the Main Injector to preserve small proton and antiproton emittances during transfer between machines. After some operational problems immediately after the shutdown, these improvements, coupled with increased reliability due to improved quench protection hardware and operating modes for the beam abort kickers, led to an increase in integrated luminosity rates, approximately 80% higher than that before the shutdown.  In early 2004, six $L_0^{max}$ and six $\int L\, dt$ per store records were set, raising these records from 48.7 to 67.5 x $10^{30}$ cm$^{-2}$ sec$^{-1}$ and from 1.9 to 3.4 pb$^{-1}$ per store, respectively.  Also records were set for $\int L\, dt$ per week = 13 pb$^{-1}$ and $\int L\, dt$ per month = 41 pb$^{-1}$.  The integrated luminosity is above the projected performance curves for FY04. A major factor in this higher performance is improved systems reliability, with fewer quenches and trips leading to longer stores, which allows more antiprotons to be accumulated, both factors leading to increased integrated luminosity rates.  A store length of ~25 hours is about optimal for the current capacity to accumulate antiprotons.  The Fall 2003 improvements have increased the ratio of luminosity per accumulated antiproton by 25-30%.  Balanced with scheduled studies required to support the upgrades and to improve performance, operations have been optimized to provide integrated luminosity.  Opportunistic machine studies are performed when appropriate, such as Booster, Main Injector, or Antiproton Source studies while the Tevatron is off for repairs, or Tevatron studies with proton beams while accumulating sufficient antiprotons for collisions.  Some of the operational problems in early 2004 included abort and chromaticity problems, and especially problems with the uncontrolled mis-steering of the beams.  There were three such occurrences which produced major quenching, resulting in vacuum leaks in the Tevatron magnets requiring 10 day long shutdowns each for component replacement, including the mid-March 2004 shutdown.  Studies are now underway to implement improved detection of beam losses and to more cleanly and safely abort the beam.

**Run II Upgrade Plans**

During 2004-2007, the Collider complex will be upgraded [2] to increase the amount of luminosity delivered per year.  The technical elements include increasing the antiproton production rate by slip stacking two proton bunches in the Main Injector, increasing the accumulation and cooling rate for antiprotons by increasing transfer line and Debuncher ring apertures, improving the Debuncher cooling and the Accumulator ring Stacktail cooling systems, completing the commissioning of the Recycler ring, and installing and commissioning electron cooling in the Recycler ring.  Beam-beam interactions in the

Tevatron will be decreased using electron beam lenses and improved helical orbits with increased proton-antiproton separations. Reliability and maintenance will continue to be addressed for the 32-year-old Linac and Booster and the 20-year-old Tevatron.

**Luminosity Projections**

These upgrade elements are technically challenging and some entail significant Research and Development. This leads to both technical and schedule risk. However, most of the upgrade elements proceed in parallel. Both the performance parameters and schedules are largely independent, except for coupling through operations and shutdowns. A Luminosity projection model has been formulated which incorporates performance parameters from parametric modeling and from data, long term operating experience, a schedule of shutdowns which includes 1-week turn-on and a performance recovery curve, phased implementation of upgrades and learning curves, and an explicit investment (the "pbar tax") of antiprotons for commissioning of the Recycler and Electron Cooling. A conservative 85 hours per week is assumed for integrating luminosity. This leads to two projections for integrated luminosity. The Design Projection does not assume the use of schedule contingency or slippage, but maintains a reasonable engineering design margin for each sub-component of the upgrade plan. The Base Projection assumes a model for schedule slippage and under-performance for all subprojects, see table below.

| Parameter | Design | Base | units |
|---|---|---|---|
| Schedule Slippage | none | 3 | months for slip stacking |
|  |  | 6 | months for other phases |
| Debuncher & AP 2 Apertures | 32 | 25 | $\pi$-mm-mrad |
| Slip Stacking | 8 | 7 | x $10^{12}$ protons per pulse |
| Average Stacking Rate | 39 | 24 | x $10^{10}$ antiprotons/hour |
| Stack in Accumulator Ring | 20 | 24 | x $10^{10}$ antiprotons |
| Stack in Recycler Ring | 570 | 360 | x $10^{10}$ antiprotons |
| Luminosity (per year) | 2.0 | 1.1 | fb$^{-1}$/year |
| Total Run II Integrated Luminosity | 8.5 | 4.4 | fb$^{-1}$ through FY 2009 |

The ultimate integrated luminosity through August 2009 would be 8.5 fb$^{-1}$ for the Design projection and 4.4 fb$^{-1}$ for the Base projection. The two luminosity projection scenarios, as functions of time, are illustrated in the two figures below.

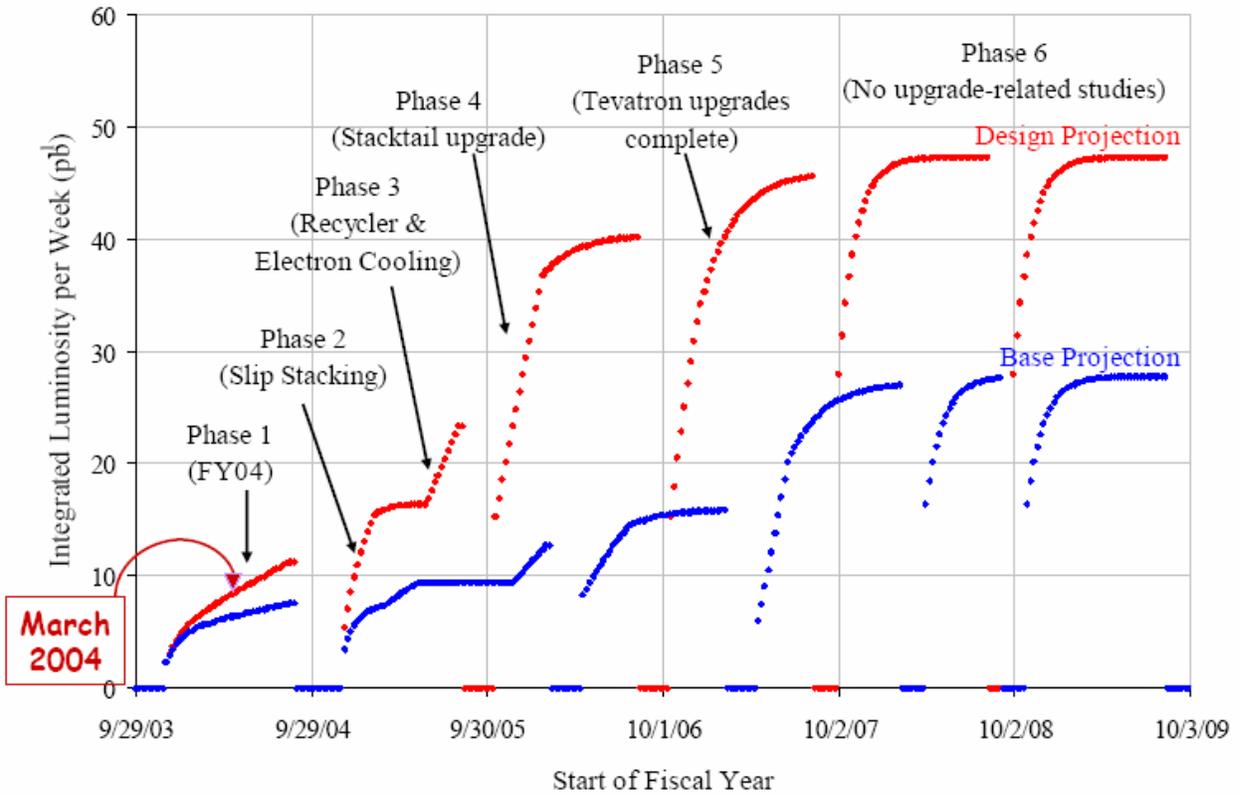

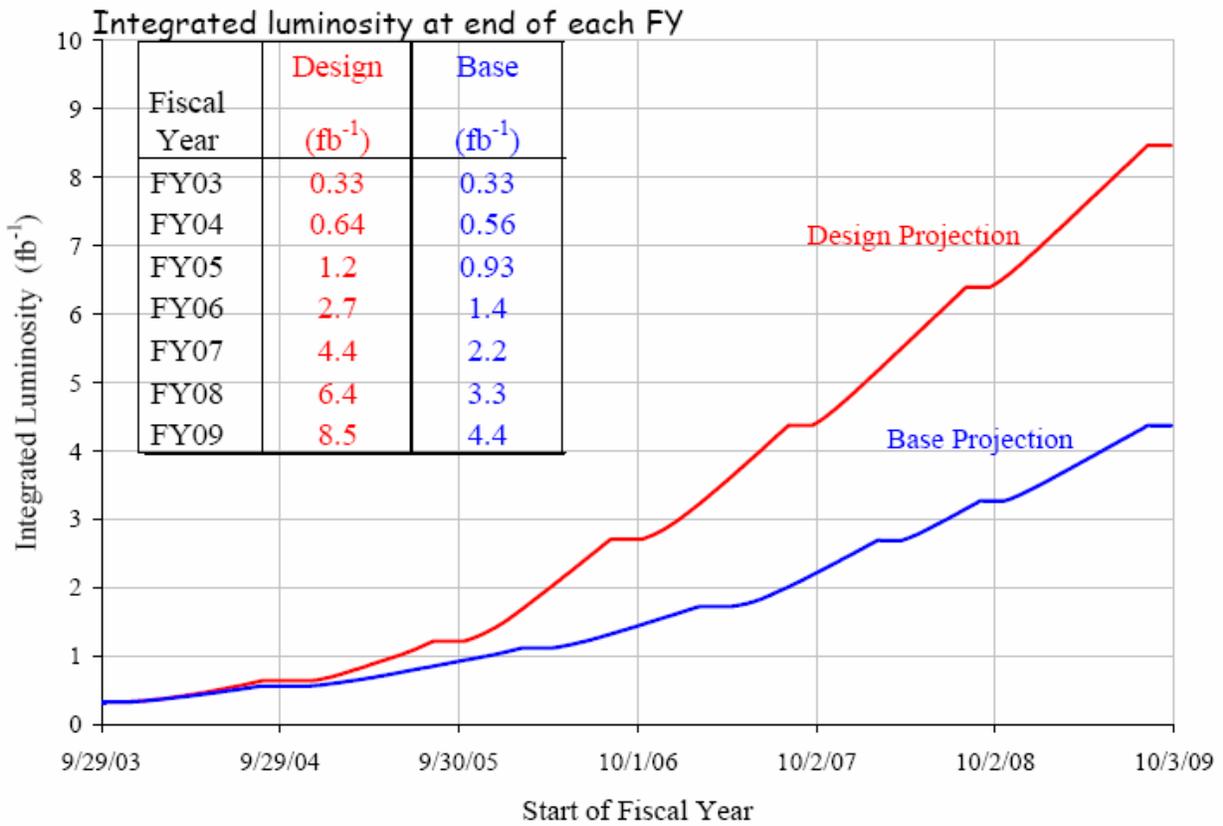

The first figure above shows the planned Tevatron shutdowns, with a model for luminosity rate recovery after the shutdown, and the implementation of individual upgrade sub-projects, with learning curves.

The Run II Luminosity Upgrade plan was reviewed and assessed by a Department of Energy panel of 16 experts in February 2004 [4]. The panel commented that:

 "the Design Projection assumes timely success with the stack-tail cooling upgrade, antiproton stacking rates, beam-beam performance in the Tevatron, and electron cooling in the Recycler...there is still significant uncertainty;

the Base Projection could be reached without the stack-tail cooling upgrade or electron cooling if there is no schedule slip in other upgrade activities and if their performance goals are met;

the committee now views the Base goal of 4.4 fb$^{-1}$ by the end of fiscal year 2009 as having a good probability of being met or even exceeded; and

meeting the Design goal of 8.5 fb$^{-1}$ by the end of fiscal year 2009 remains a very challenging goal."

**Acknowledgements**

This report represents a lot of hard work by many members of the Fermilab Accelerator, Technical, Computing, and Particle Physics Divisions. I especially wish to thank the Run II Upgrade project leaders Jeff Spalding and Dave McGinnis whose prior presentations were used as my starting point.

**References:**

1. Run II Handbook, 2001     http://www-ad.fnal.gov/runII/index.html
2. The Run II Luminosity Upgrade at the Fermilab Tevatron, v2.0, January, 2004
        http://www-bdnew.fnal.gov/doereview04/RunII_Upgrade_Plan_v2.0.pdf
    individual upgrade element presentations from the February 2004 review can be found at     http://www-bdnew.fnal.gov/doereview04/index.htm
    also see the earlier plan v1.0 June 2003 for more details
        http://www-bd.fnal.gov/doereview03/docs/Overview7.1.pdf
3. Luminosity plot is updated weekly
        http://www.fnal.gov/pub/now/tevlum.html
4. DOE review panel closeout – February 2004
    http://www-bdnew.fnal.gov/doereview04/RunII_Lum_Review0204_closeout.pdf